\documentstyle[twocolumn,aps]{revtex}

\begin{document}
\title{Modeling Market Mechanism with Evolutionary Games}
\author{Y.-C. Zhang\\{\small Institut de Physique 
Th\'eorique, Universit\'e de Fribourg, 1700 Fribourg, Switzerland}}
\maketitle

{\bf Recent Trend}

A new trend arises that more and more physicists are attracted to economy-related
problems. This is evidenced by the growing numbers of publications in physics journals
devoted to theoretical and applied issues in economics and finance, fresh Ph.D's and 
seasoned researchers alike find career on Wall Street, and new journals are launched and
conferences are organized etc. Trespassing on others' domain is notorious for physicists:
their insatiable curiosity steadily pushes them into their near and far neighbors like
biology, economics and other natural and social disciplines. 

However, the current predilection seems to be in economy, especially in finance. Probably one
can recognize two categories of interest: 1) physicists' tools are much in demand and the so-called
"technical analysis" is more and more complex and often an experienced physicist can offer the much
needed skills that traditionally trained economists lack; 2) research topics present fundamental,
intellectual challenges, the aim is to understand the basic mechanism. In this essay we shall concentrate
on this second category. Moreover, we shall limit ourselves to modeling market mechanism.

{\bf Why Interesting?}

To appreciate why economy is interesting for physicists and what challenge is left there one has to know
what is the current accepted theory in that field. We must not pretend to summarize the state of art
here, a critical appraisal can be found in a recent {\it Santa Fe} Proceedings [1]. If we look into the
literature, as a physicist one may get the strange feeling that the Theory is so detached from the Experiments.
On the one hand the Theory is extremely refined and self-consistent with little effort to compare with empirical
evidence; on the other the experiments (e.g. traders' activities) are also extensively performed with little
reference to the Theory. It is revealing to see how Soros, one of the top players in finance considers the Theory [2].
In short, current prevailing theory assumes equilibrium, the descriptions are mostly static. Little is said how
an equilibrium can be attained, if attainable at all. One get impression that the Theory looks like "mean field"
theory in physics, balancing actions and reactions. Dynamics of markets can be found nowhere. Meanwhile, Insight 
gained in the statistical mechanics, especially in 
non-equilibrium processes, may inspire physicists to have a try in formulating a sort of dynamical theory for some
economical processes.

Experimental activities in finance are highly developed, much data (e.g. crashes) defy explanation. One is tempted to 
compare the current state of affairs to the thermodynamics before Boltzmann or even Carnot, where the framework is
yet to be established. One does not know how to put together pieces of empirical law into a coherent picture. Let us
not entertain this comparison too far, since economy provides less precise data and the fundamental elements are
{\it thinking agents} versus the obedient particles in thermodynamics. One can never hope to get a future economy
theory as quantitative and predictive as physical laws. However, this should not deter us from searching a
framework to understand some basic phenomena qualitatively. In short, any relevant theory should remain in "soft science",
we should keep this in mind when modeling economy. 

Studying economy related problems can give certainly valuable feedback for physics proper. 
Barring the risk of offending our colleagues, we might say that current statistical mechanics research is somewhat stagnant, 
respect to its exciting days in the 70's and 80's. Economy presents new challenges, 
new problems and new ways of thinking, can teach us some new secrets of how Nature works in this "soft" science discipline.

{\bf How to Put Hands on?}

Lacking a general framework, one has to search models in the dark, thus various models by physicists appeared in the literature.
To start one needs a concrete model. Before present our own model let us recall that our aim has to be very modest, the best one can
have is a sort of paradigm. One has also to keep the model as simple as possible, in order to say something general. One has
here in mind the Ising Model, despite its oversimplification, still offers insight into real magnetism; or the BTW 
Self-Organized Criticality Model [3], supposedly applies only to ideal sand piles, but offer insight into many natural
phenomena. Let us list a minimal set of ingredients that are indispensable for modeling markets.

1) A large number of independent agents participate in a market.   
 
2) Each agent has some alternatives in making his decision.

3) The aggregate activity results in a market price, which is known to all.

4) Agents use the public price history to make their decisions.

This set of ingredients is quite arbitrary, we omit notably two important factors: a) no fundamental
news in the market besides their own trading activity; b) agents do not believe in any theory, they only
learn from their own experience and believe, that the price history contains information. We want to study the 
inherent dynamics of market, in the absence of external influences. Real economy has both internal and external
contributes. B. Arthur has advocated the so-called "inductive thinking" approach [4], which represents a minority
opinion in economics. His idea is that since an agent cannot use the Theory to deduct his decision, his only choice is
to learn from his own experience, as many a trader would attest. Our own model is inspired from Arthur's {\it El Farol} problem [4]. 
 Below we shall illustrate our ideas through two models. The first Competition
Model is intended to reveal the rich intrinsic market dynamics and general issues; the second Trading Model is to
apply the basic ideas to an artificial market.

{\bf The Minority Model} 

The simplest model we can think of is defined in a form of evolutionary game.  
Let us consider a population of $N$ (odd) players. 
At each time step, everybody has to
choose to be on side $A$ or $B$. The payoff of the game is
that after everybody has chosen side independently, those who
are in the {\bf minority} side win. In the simplest version, all winners
collect a point. The players make decisions based on the common
knowledge of the past record. The past record 
only tells which side is winner, without
the actual attendance number. Thus the time series can be represented by a
binary sequence, 1 or 0 meaning $A$ or $B$ is the winning side.

Let us assume that our players are quite limited in their analyzing
power, they can only retain last $M$ bits of the system's signal and
make their next decision basing only on these $M$ bits. 
Each player
has a finite set S of strategies. A strategy is defined to be the next action (to
be in $A$,$B$, or 1,0), given a specific signal's M bits. The example of one
strategy is illustrated in table 1 for M=3.

\begin{center}
\begin{tabular}{|c|c|}\hline
        record&decision\\ \hline
        000&1\\ \hline
        001&0\\ \hline
        010&0\\ \hline
        011&1\\ \hline
        100&1\\ \hline
        101&0\\ \hline
        110&1\\ \hline
        111&0\\ \hline
\end{tabular}
\end{center}

There are 8 ($=2^M$) bits we can assign to the right side. Each configuration
corresponds a distinct strategy, this makes the total number of
strategies to be $2^{2^M}=256$. This is indeed a fast increasing number,
for $M=2$, $3$, $4$, $5$ it is $16$, $256$, $65536$, $65536^2$. We
randomly draw $S$ strategies for each player, from the pool.
All the $S$ strategies in a player's bag can
collect points depending if they would win or not given the $M$ past
bits, and the actual outcome of the next play. However, these
are only $virtual$ points as they record the merit of a strategy as if it
were used each time. The player uses the strategy having the highest
accumulated points for his action, he gets a real point only
if the strategy used happens to win in the next play. The fact of using alternative strategies  makes the players
adaptive to the market. A player thus tends to maximize his capital (cumulated points) and his performance
is judged only on his time averaged capital gain.

Several remarks are in order. By the very definition of {\bf minority} agents are not
encouraged to form commonly-agreed views on the market. This is like in real markets
bears and bulls live together. In real trading it is often observed that a minority of traders get into a trend (buying or selling)
first, the majority get finally dragged in also. When the minority anticipates correctly and get out of the trend in time, they 
pocket the profit at the expense of the majority.  There is limited resource available for competition.
If the players manage to coordinate well, {\it per} play they can expect $(N-1)/2$ points, the maximal
gain possible. Since our players are selfish, no explicit coordination is imposed, their fate
is left to the market. The important question is if they can somehow learn to spontaneously 
cooperate. $S=1$ simplifies further the model in which instead of players the strategies compete directly.  The 
outcome is trivial deterministic signal. It is the extra layer of complexity at the player's level
ensures adaptability. (Remember the role of hidden layers in Neural Network?)

This binary model is very suitable for numerical experiments. Damien Challet, a graduate student
at Fribourg, implemented the game on a computer. The preliminary results are reported in
[5]. Note the word "experiment" instead of "simulation" is used. This is to emphasize that we did not have precise
goals at the start,
like in an exploratory experiment, the players are let
loose to play and we observe. We were rather amazed with the complex, rich consequences of the model.

The parameters of the model are just three: $N,M,S$. However, there hidden parameters. The space of the
strategies is most intriguing. Apparently this space is so huge, that for realistic parameter values say
$M=10$, this space should be regarded as infinite ($
2^{2^M} > 10^{300}$) for all practical purposes. Numerical experiments show that the space is far from
infinite. Depending on $N$ the market has distinct behavior, $M=10$ can be too large or too small for achieving coordination.
How can such large number ($10^{300}$) still be relevant in this model? A more refined analysis of the strategy space 
can solve this paradox of large number. 

{\bf Boolean "Genetic" Space}

It is instructive to represent the strategy space on a $2^M$ dimensional Boolean hypercube. The $N_{tot}=2^{2^M}$ distinct
strategies are on the points of the hypercube. There is a striking similarity with Kauffman's Boolean $NK$ network [6]. 
Consider two neighboring strategies which differ only by one bit,
in other words the Hamming distance between them is one. These two strategies predict almost always the same outcome acting 
on the past record: out of $2^M$ possibilities only one exception. Therefore the distinct strategies
are highly correlated. Players using correlated strategies tend to obtain the same decision, thus hindering their chance
finding the minority side. Among the $N_{tot}$ strategies there is huge redundancy. If two strategies are uncorrelated, 
their decision outcomes should match with 1/2 probability.
This is possible if their Hamming distance is 1/2 of the maximal value ($2^M$).
We are thus led to count
mutually uncorrelated strategies. This count will provide a crucial measure
of diversity (or independence) of the $N_{tot}$ strategies. 

We recall some useful basic properties of a hypercube: There is a subset of
$2^M$ pairs of points (out of $N_{tot}$, within every pair the Hamming distance is maximal ($2^M$). They are anti-correlation pairs in the sense
two strategies of a pair always predict the opposite actions. All other Hamming distances among these $N_0=2^{M+1}$ points
are $2^{M-1}$, i.e. mutually independent. In other words these $N_0$ points are composed of two groups of $2^M$ points each.
Within the group the strategies are completely independent. Some of the cross-group links can be anti-correlated.  

It is this reduced number $N_0$ which plays an important role in the model. If the number of strategies in the population 
$NS$ is larger than
$N_0$, then the players have to use strategies which are positively correlated, the herd effect is unavoidable despite the
adaptability of our players. The herd effect will result in fluctuations larger than random chance would warrant, thus 
leads to waste of the limited resources.On the other hand, if $NS<<N_0$,
the "independent" subset of strategies indeed appears to be independent. $NS$ cannot sample enough the $N_0$ strategies,
the anti-correlation Hamming distances can hardly be represented (correction being about $NS/N_0$). 
The players in this case will appear as if they were using
random strategy, independent from each other. Most interesting is the critical case when $NS\sim N_0$, i.e. when the
reduced set is more or less covered by the population. The majority of their mutual Hamming distances implies independence,
but a small part (about square-root of the total) implies anti-correlation. This means that the players 
behave almost independently,
the small number of anti-correlation Hamming distances help them to obtain opposite  decisions. This is beneficial for everybody
since coordinated avoidance makes the players more evenly distributed on the two sides, the limited 
resource is better exploited.   
Therefore there are three distinct phases: 1) the overcrowded phase $NS>N_0$ where positive correlation is inevitable, 
the herd effect 
makes
it worse for everybody; 2) a cooperation or critical region $NS\sim N_0$ where the strategies used by $N$ players are mostly 
independent, plus some anti-correlation links to achieve mutual avoidance, the resource is better shared; 3) a random region
$NS<<N_0$ where the anti-correlation is hardly present, strategies appear to be independent. A recent numerical
study [7] is in qualitative agreement with the above  analytical result.

{\bf Darwinism and Evolution towards cooperation}

Between the over-competitive and random phases lies the critical region $NS\sim N_0$ where cooperation emerges. This reminds us
of Kauffman's  often quoted motto "adaptation toward the edge of chaos".
The competing population has different phases of collective behavior, depending on the choice of $N,M,S$. We see from the above
discussion that adaptation puts an {\it evolutionary pressure} on the players' strategies, they tend to be as diversified as 
possible. In the $NS>N_0$ region, this pressure (or tendency) is frustrated since the strategy space (the hypercube) is too
crowded. To relieve this pressure, one may want to change $S$ or $M$ for fixed $N$. Changing $S$ is not very efficient. 
Increasing $M$ is highly effective because of the exponential dependence.

We would like to let the population to evolve to
this cooperative region without imposing the fixed parameters. To this end we introduce another level of adaptation by adding 
the Darwinism to this competition game. Worst player is periodically wed out of the game; best player is allowed to produce an
offspring. The new born player is a clone copy of his parent with the virtual capital reset to zero. A small mutation rate is 
assumed to ensure genetic diversity, i.e. one of his inherited strategies is replaced. The total population is kept constant
without loss of generality. The newly drawn strategy can have its $M$ changed by one unit. This permits the players to find 
the best suitable $M$. The population can nevertheless reach a stationary state, clearly now with inhomogeneous $M$. In short, 
the population is able to evolve without outside intervention and  self-organize 
themselves to find the critical region $NS\sim N_0$,
where it is beneficial for everybody.[5]  

Another interesting variant is to let the strategies fight directly, without players as intermediary ($S=1$). In order to have
nontrivial dynamics going on, we allow a large number ($N_1$) of randomly drawn strategies to participate the game passively,
only a small number of strategies of the total $N_a<<N_1$ actively play. The passive strategies observe the system's posting board
and count their virtual gain, i.e. as if they took part in the game. If one strategy's simulated gain is above a 
certain threshold, say
any benchmark average in the actual game, he is bullish enough to come to the fore by becoming an active strategy. The worst 
strategies  recede to the status of passive observers. Each strategy has two balance sheets, one for the cumulated real gain
during his active playing periods; another for simulated gain counted from the last time he is relegated to the observer pool.
In this variant the Darwinism evolutionary pressure is also operative. Positively correlated strategies do well as observers,
but do poorly in actual play. Therefore their active appearance is bound to be brief. Most interesting is to let the number
of active strategies $N_a$ free, everyone is allowed to actively play, the market should decide how many actually do. Clearly inhomogeneous
$M$ is better and mutation should also be operative in order to keep an adequate diversity. The system in a way is like a 
grand-canonical ensemble. 
A large pool of observers is also a realistic feature
of markets. Many an investor figures out his chances by simulation, he is attracted to markets if he believes that he has
an edge. But often the subsequent participation can be disappointing. 
 
{\bf A Prototype Trading Model}

The minority model, though very rich, still lacks some most basic features in a real market, e.g. the price which is determined
by the aggregate supply and demand. We want to keep the above general ingredients, to build  a more specific trading model.
In a trading model players have to decide when to buy and sell, just like to take $A$ and $B$ side in the minority model.
However the payoff is not a fixed rule, but depends on the price movement which is in turn determined by the players'
trading. 
In the modern market of stocks, currencies, and  commodities, trading patterns 
are becoming more and more global. Market-moving information is available to
everybody. However, not  all  the participants
interpret the information the same way  and  react at the same time delay. In
fact, every participant  has a certain fixed framework facing external events.
It is  well  known that the global market is far from being at  equilibrium [1],
the collective behavior of the  market can  occasionally have violent bursts.

Let us define our model more precisely. Each player is 
initially given the same amount of capital in two forms: 
cash and stock. There is only one stock in this 
model.
All trading consists of switching back and forth between 
cash and this stock. Each player has a strategy that 
makes 
recommendation for buying or selling a certain amount of 
stock for the 
next time step. This depends solely on the price history.
Player $i$'s strategy is an arbitrary  nonlinear function 
$F_i (p_t,p_{t-1},\ldots )$, positive (negative)
values recommend the amount of buying (selling). The aggregate trading 
decides the next price tick $p_{t+1}$, using the law
of supply-demand. Darwinism is also implemented here.

Our numerical simulations are quite encouraging [8]. Despite the 
simplicity and the arbitrariness of the strategies, 
an extremely rich price history is created. 
A sample of $p_t$ is shown in fig.1,
which shows fluctuations of all sizes. 
Depending on the parameters, on long 
runs, also some crushes occasionally occur, with no warning. 
New features appear here respect to the minority model: Even though the same
self-organized structure is used here, the system does not appear to reach equilibrium,
one can hardly limit the range of the price fluctuations. This is due to partly the continuous
strategy space, as well as the law of supply-demand.

{\bf Open Questions and Perspectives}

We have discussed two models of self-organized systems, in which players compete in a common market place using 
the signal produced by their own activities. We argue that this general scenario should be also present in real markets.
As in an early stage of any emergent science discipline (some already baptized "econophysics"), many different
models are proposed. We feel the current need is learn how to ask the right questions about economical processes.
By asking right questions and trying to answer them, we have to explore many seemingly isolated models, empirical laws
etc to be able to set up a workable framework. Already at the level of simplest minority model, we see that many 
interdisciplinary
subjects have naturally been related to. Let us indulge ourselves with an arbitrary list: self-organized criticality, population
and Darwinism, ecology, information science (auto-feeding signals should be analyzed a la Shannon), glasses and
spin-glasses (frustrations, non-ergodicity), Kauffman's NK model and auto-catalysis, game theory (Prisoner's dilemma pits 
two players to compete, the minority game can be a natural generalization) etc. Many of these relations deserve
further study. Besides some relevance to economy, continuing this  exercise of model building and playing is 
certainly rewarding. 

Acknowledgments: I thank D. Challet, G. Caldarelli and M. Marsili for their fruitful collaboration.

Figure Caption: Price history is the artificial trading model. Self-similarity seems to suggest some scaling invariance,
in fact the statistical properties is in reasonable agreement with real market data [8].

{\bf References :} 

[1] {\em The Economy as an Evolving Complex System}, 
ed. P.W. Anderson, K. Arrow and D. Pines, Redwood City, Addison-Wesley
Co.,1988

[2] G. Soros, {\em Alchemy of Finance}, J. Wiley \& Sons Inc., New York, 1994.

[3] P. Bak, C. Tang, and K. Wiesenfeld {\em Self-Organized Criticality}, 
Phys. Rev. Lett. 59 (1989) 381.

[4] W. B. Arthur, {\em Inductive Reasoning and Bounded Rationality}, Am. Econ. Assoc. Papers and Proc. 84, pp. 406-411, 1994

[5] D. Challet and Y.-C. Zhang, 
{\em Emergence of Cooperation and Organization in an Evolutionary Game}, 
Physica A 246 (1997) 407 
(also at adap-org/9708006); and  unpublished 1998.

[6] S. Kauffman, {\em The Origins of Order}, Oxford Univ. Press, New York 1993.

[7] R. Savit, R. Manuca and R. Riolo, Michigan Univ. Preprint, {\em
Adaptive Competition, Market Efficiency, Phase Transitions and
Spin-Glasses}, 
adap-org/9712006, Dec. 1997.

[8] G. Caldarelli, M. Marsili and Y.-C. Zhang, {\em A prototype model of stock
exchange}, Europhys. Lett. 40 (1997) 479

\end{document}